\documentclass[aps,prd,amsfonts,amssymb,preprint,eqsecnum,nofootinbib,superscriptaddress]
{revtex4} 
\usepackage{graphics,epsfig}

\newcommand{\pslash}{p\llap{/\kern-0.3pt}}
\newcommand{\qslash}{q\llap{/\kern-0.3pt}}
\newcommand{\rslash}{r\llap{/\kern-0.3pt}}
\newcommand{\lslash}{\ell\llap{/\kern-0.3pt}}
\newcommand{\lsim}{\mathrel{\rlap{\lower4pt\hbox{\hskip1pt$\sim$}}
    \raise1pt\hbox{$<$}}}                % less than or approx. symbol
\newcommand{\gsim}{\mathrel{\rlap{\lower4pt\hbox{\hskip1pt$\sim$}}
    \raise1pt\hbox{$>$}}}                % greater than or approx. symbol
\begin{document}
\preprint{WM-06-113}
%
% Title of paper
\title{\vspace*{0.5in} Holographic Bosonic Technicolor
\vskip 0.1in}
\author{Christopher D. Carone}\email[]{cdcaro@wm.edu}
\author{Joshua Erlich}\email[]{jxerli@wm.edu}
\author{Jong Anly Tan}\email[]{jmtanx@wm.edu}
\affiliation{Particle Theory Group, Department of Physics,
College of William and Mary, Williamsburg, VA 23187-8795}
\date{December 2006}
\begin{abstract}
We consider a technicolor model in which the expectation value of an
additional, possibly composite, scalar
field is responsible for the generation of fermion masses.  We define the
dynamics of the strongly coupled sector by constructing its holographic dual. Using the AdS/CFT
correspondence, we study the $S$ parameter and the phenomenology of the light technihadrons.
We find that the $S$ parameter is small over a significant region of the model's parameter space.
The particle spectrum is distinctive and includes a nonstandard Higgs boson as
well as heavier hadronic resonances.  Technihadron masses and decay rates are calculated
holographically, as a function of the model's parameters.
\end{abstract}
\maketitle

\section{Introduction} \label{sec:intro}

The physics responsible for the breaking of electroweak symmetry will
be studied at the LHC over the next few years.  In anticipation of
potentially definitive experimental results, a number of novel
models of electroweak
symmetry breaking (EWSB) have been proposed recently
\cite{EWSB-models}.  These models aim, in part, to address
the hierarchy problem that is inherent to the Higgs sector of the minimal Standard Model,
while satisfying the constraints posed by LEP data.  Long ago, technicolor models were proposed as
an alternative to the minimal Higgs sector \cite{technicolor}.  Fermions coupling both to
the electroweak and technicolor gauge sectors condense when the technicolor interactions
become strong. The fermion condensate takes the place of the vacuum expectation value (vev)
of the Higgs field in the breaking of electroweak symmetry.  The hierarchy problem associated
with radiative corrections to the Higgs mass is eliminated since no fundamental scalar
fields are present in the theory.  The large separation between the Planck and electroweak
scales is understood as a natural consequence of the logarithmic running of the
technicolor gauge coupling. Unfortunately, it was realized in the last millennium that
technicolor models predict large corrections to precisely measured electroweak observables
if the new strong sector is similar to QCD~\cite{PT}.   In particular, the Peskin-Takeuchi
oblique parameter $S$ is predicted to be $\agt 0.2$, while experiments constrain $S$ to be less
than about $0.1$~\cite{rpp}.

More recently, technicolor models with small or negative values of $S$ have been constructed
using the AdS/CFT correspondence~\cite{sanz,hong,piai,Agashe}.   In this holographic approach, the dynamics of
the strongly coupled sector is not necessarily specified {\em a priori}. Instead,  a
five-dimensional (5D) gauge theory in a warped background is postulated to define a strongly
coupled four-dimensional (4D) technicolor sector via the rules of the AdS/CFT
correspondence~\cite{AdSCFT}.  Beginning with a holographic theory that has properties similar
to QCD, new couplings are introduced in the 5D model that alter the holographic prediction of
current-current correlation functions in the strongly-coupled theory.  It can then be shown
that the parameter space of the 5D theory contains regions where the value of the $S$ parameter
is in accord with experimental constraints~\cite{sanz}. In the present work, we also use the
freedom to deviate from a QCD-like holographic theory in constructing a viable model of
dynamical EWSB.  In particular, we will allow for a separation of the technicolor
confinement and chiral symmetry breaking scales as another means for reducing the $S$-parameter.

A complete model of holographic technicolor must also provide a
mechanism for generating Standard Model fermion masses.  Four-dimensional extended
technicolor (ETC) models provide the desired coupling between
Standard Model fermions and
the technicolor condensate through four-fermion operators that are generated when heavy ETC gauge
bosons are integrated out of the theory~\cite{etc}.  Unfortunately, ETC gauge boson exchange generically
produces flavor-changing four-fermion operators that are excluded by experiment, if the ETC
scale is low enough to account for a heavy top quark.  An alternative means of generating
fermion masses is possible in technicolor theories that have an additional,
possibly composite, Higgs doublet field in the low-energy theory~\cite{kag,sim,cg,cs1,cgold,cs2,others,ss}.
The coupling of the technifermion condensate to this field forces it to develop a vev, even if the
Higgs mass squared is positive.  Yukawa couplings of the Higgs field then provide the origin of
Standard Model fermion masses. If the Higgs field is composite, we assume that
the compositeness scale is higher than the technicolor scale, so that the Higgs may be treated as a new
fundamental scalar in the low-energy effective theory. While it may seem unusual to consider
electroweak symmetry breaking models with fundamental scalars, viable strongly coupled extended
technicolor sectors that suppress flavor changing operators often have scalars in their effective
description below the ETC scale \cite{setc}. The basic features of this ``bosonic
technicolor" scenario are reviewed in Section~\ref{sec:4DModel}.

In this paper, we present a bosonic technicolor model in which the dynamics of
the strongly coupled sector is defined through its holographic dual.  The model
is compatible with electroweak constraints and provides for the origin of fermion masses.
Coefficients in the electroweak chiral Lagrangian of the theory, that would otherwise be unknown,
are determined by the AdS/CFT correspondence, as we discuss in Section~\ref{sec:AdSCFT}.  We then study
the phenomenology of the model in Section~\ref{sec:Pheno}. In particular, we compute the usually problematic
contribution to the $S$ parameter, as a function of the technirho mass and the vev of the Higgs
field, and find that a significant region of parameter space is allowed.  We also study the decays of the
technirho which, if observed at the LHC, could exclude regions of the model's parameter space and
potentially discriminate between different holographic technicolor scenarios. Neither the $S$ parameter
nor the partial decay widths of the technicolor resonances were calculable in earlier
versions of bosonic technicolor. We conclude in Section~\ref{sec:Conclusions}.

\section{Vintage Technicolor with a Scalar}\label{sec:4DModel}
The gauge group of the model is $G_{TC} \times$SU(3)$_C\times$ SU(2)$_W\times$U(1)$_Y$, where
$G_{TC}$ represents the technicolor group.  We
will assume
that $G_{TC}$ is asymptotically free and confining, but make no other assumptions about the group.  We assume two
flavors of technifermions, $p$ and $m$, which transform in a nontrivial
representation of $G_{TC}$.
In addition, these fields form a left-handed SU(2)$_W$ doublet and two right-handed singlets
\begin{equation}
\Upsilon_L \equiv \left(\begin{array}{c} p\\m \end{array} \right)_L \, , \,\,\,\,p_R\,,\,\,\,\,m_R,
\end{equation}
with hypercharges $Y(\Upsilon_L)=0$, $Y(p_R)=1/2$, and $Y(m_R)=-1/2$. With these assignments, the
technicolor sector is free of gauge anomalies.

The technicolor sector has a global SU(2)$_L\times$SU(2)$_R$ symmetry, which corresponds
to independent special unitary rotations on the left- and right-handed technifermion fields.  It is assumed
that strong dynamics results in a technifermion condensate
\begin{equation}
\langle \bar{p}p+\bar{m}m \rangle \approx 4 \pi f^3 \,\,\,
\label{eq:condef}
\end{equation}
that spontaneously breaks this symmetry.  Here, $4 \pi f$ is traditionally identified as
the chiral symmetry breaking scale~\cite{georgi}.  The resulting Goldstone bosons may be described in an
effective chiral lagrangian, where
\begin{equation}
\Sigma = \exp(2 \, i\, \Pi /f) \,\,\,\,\mbox{ and } \,\,\,\,\,\,
\Pi = \left(\begin{array}{cc} \pi^0/2 & \pi^+/\sqrt{2} \\ \pi^-/\sqrt{2} & -\pi^0/2 \end{array} \right),
\label{eq:sigpi}
\end{equation}
and where the $\Sigma$ field transforms simply under the SU(2)$_L\times$SU(2)$_R$ symmetry:
\begin{equation}
\Sigma \rightarrow L \, \Sigma\, R^\dagger.
\label{eq:sigtrans}
\end{equation}
A kinetic term for $\Sigma$ may be constructed that is invariant under Eq.~(\ref{eq:sigtrans}), and
also under the Standard Model gauge symmetries,
\begin{equation}
{\cal L}_{{\rm KE}} = \frac{f^2}{4} \mbox{Tr }(D_\mu \Sigma^\dagger D^\mu \Sigma) \,\,\,,
\label{eq:kin1}
\end{equation}
where the covariant derivative is given by
\begin{equation}
D^\mu\Sigma = \partial^\mu \Sigma - i g W^\mu_a \, T^a \Sigma + i g' B^\mu \Sigma \, T^3.
\end{equation}
Here, the $T^a$ are the generators of SU(2), while $g$ and $g'$ are the SU(2) and U(1) gauge couplings,
respectively.  The quadratic terms in Eq.~(\ref{eq:kin1}) include mixing between gauge fields and
the pion fields, indicating that the latter are unphysical and can be gauged away.  After doing so,
the remaining quadratic terms in Eq.~(\ref{eq:kin1}) give the gauge boson masses,
\begin{equation}
m_W = \frac{1}{2}\, g f \,\,\,\,\,\,\,\,\,\, m_Z =\frac{1}{2}\,(g^2 +{g'}^2)^{1/2} f \,\,\,,
\label{eq:gbmass}
\end{equation}
which reproduce the correct experimental values for $f \approx 246$~GeV.

For $G_{TC}=SU(N)$, the theory described thus far corresponds to conventional technicolor, with all
its well-known problems.  Large contributions to the $S$ parameter will be avoided in our case
by deforming the model away from one that could be naively interpreted as a scaled-up version of QCD.  This will
be discussed in the next section.  Here, we extend the model to provide for an origin of fermion masses.  We
assume that the low-energy theory includes a scalar SU(2)$_W$ doublet $\phi \equiv (\phi^+ , \phi^0)$ that
can couple to the technifermions and to Standard Model fermions via ordinary Yukawa couplings:
\begin{equation}
{\cal L}_{\phi\, T} = \overline{\Upsilon}_L \tilde{\phi} h_+ p_R +
\overline{\Upsilon}_L \phi h_- m_R + \mbox{ h.c.}  \,\,\,,
\end{equation}
\begin{equation}
{\cal L}_{\phi\, f} = \overline{L}_L \phi h_l E_R +
\overline{Q}_L \tilde{\phi} h_U U_R+
\overline{Q}_L \phi h_D D_R+\mbox{ h.c.}
\end{equation}
We will {\em not} assume that $\phi$ has a negative squared mass.  The Yukawa coupling of $\phi$ to
the technifermions produces a $\phi$ tadpole term when the chiral symmetry is dynamically broken and
the technifermions condense. This guarantees that there is a non-zero vacuum expectation value for $\phi$,
and hence, that masses for the Standard Model fermions are generated.

%%%%%%%%%%%%%
The origin of the $\phi$ doublet is worthy of some comment.  The $\phi$ field either represents a fundamental
particle in the ultraviolet (UV), or a composite one in the infrared (IR).  Each possibility presents its
own advantages and disadvantages. If $\phi$ is fundamental, then an ultraviolet completion that separately solves
the hierarchy problem, such as supersymmetry, must be assumed.  Some may object to such a hybrid proposal on
philosophical grounds, but such arguments have little bearing on whether or not such a theory is realized in
nature.  On the other hand, if $\phi$ is composite, one avoids the problems of quadratic divergences, which are
regulated by the Higgs compositeness scale.  In this case, however, other difficulties may occur.  The couplings
of the composite field to the fundamental fermions in the theory arise via higher-dimension operators, leading to
suppression factors.  One might worry that the top quark Yukawa coupling could be too small in a generic model
of this sort, though such an outcome could be avoided, for example, if the third generation fermions are also
composite.  A Higgs compositeness sector may also provide a new source for dangerous flavor-changing neutral
current effects, via higher-dimension operators suppressed by the Higgs compositeness scale.  Whether such
problems actually do arise, however, depends on the details of the theory in the UV, which are unknown.
Since we work exclusively with the low-energy theory, it is only necessary that we assume that some
adequate UV completion of our model exists.  The same assumption is made in other popular models of
EWSB~\cite{EWSB-models}.

We should also comment on naturalness of bosonic technicolor models.  We
assume that the scalar field in the effective low-energy Lagrangian
has a positive squared mass. The scalar mass cannot
be arbitrarily large, or else the scalar vacuum expectation value
would not be large enough to account for fermion masses with perturbative
Yukawa couplings.  Therefore, the scalar sector of
bosonic technicolor is comparable in naturalness to the Higgs sector
of the Standard Model.  We always assume that if bosonic technicolor is
realized in Nature, then it is the low-energy effective description of a
theory in which the scalar mass is stabilized by some additional mechanism.
We stress that our purpose is not to solve the
hierarchy problem, but to study the phenomenology of this class of
electroweak symmetry breaking models.

%%%%%%%%%%%%%
We may incorporate $\phi$ into the chiral Lagrangian by defining the matrix field
\begin{equation}
\Phi = \left(\begin{array}{cc} \overline{\phi^0} & \phi^+ \\ -\phi^- & \phi^0 \end{array}\right) \,\,\,,
\label{eq:Phi}\end{equation}
which transforms precisely in the same way as $\Sigma$ under the chiral symmetry.  For the case
in which the technifermions have a common Yukawa coupling $h_+ = h_- \equiv h$,
which we assume henceforth, the $\phi$ tadpole
described above appears through the following term in the effective chiral lagrangian~\cite{cg}
\begin{equation}
{\cal L}_H = c_1 \cdot 4 \pi h f^3\, \mbox{Tr }(\Phi \Sigma^\dagger) + \mbox{h.c.}  \,\,\,.
\label{eq:keyterm}
\end{equation}
The coefficient has been chosen such that $c_1$ is of order unity by naive dimensional analysis (at
least in QCD-like models)~\cite{nda}.  It is now convenient to employ a nonlinear representation of the $\Phi$
field,
\begin{equation}
\Phi = \frac{(\sigma+f')}{\sqrt{2}} \, \exp(2\,i \Pi'/f') \,\,\,,
\end{equation}
where $f'$ is a vev and $\Pi'$ is defined in analogy to Eq.~(\ref{eq:sigpi}). The
fields $\{\sigma, {\Pi'}^a\}$ are equivalent to the four real degrees of freedom in the original field
$\phi$. Expanding Eq.~(\ref{eq:keyterm}), one obtains the mass matrix for the $\Pi$ and $\Pi'$ multiplets.
One linear combination, which we call $\pi_a$, is exactly massless and becomes the longitudinal components
of the weak gauge bosons in unitary gauge; the orthogonal combination, $\pi_p$, are physical and remain
in the low-energy spectrum:
\begin{equation}
\pi_a = \frac{f\,\Pi+f'\,\Pi'}{\sqrt{f^2+{f'}^2}} \,\,\,, \,\,\,\,\,\,\,\,
\pi_p = \frac{-f'\,\Pi+f\,\Pi'}{\sqrt{f^2+{f'}^2}}  \,\,\,.
\label{eq:piapip}\end{equation}
Note that our phase conventions have been chosen to agree with those found in the previous
literature~\cite{cg}.  The $f$ and $f'$ vevs, as well
as the mass of the $\sigma$ field, can be determined in terms of the tadpole parameter
$c_1$ in Eq.~(\ref{eq:keyterm}), the Yukawa
coupling $h$, and
%a quartic coupling $\lambda$
the parameters that appear
in the $\phi$ potential (for a detailed treatment,
see Ref.~\cite{cg}).  For
our present purposes, we will find it more convenient to express quantities of phenomenological
interest in terms of $f$ and $f'$ directly.  The mass of the physical pion multiplet also
follows from ${\cal L}_H$ in Eq.~(\ref{eq:keyterm}).  One finds
\begin{equation}
m_\pi^2 = 8 \sqrt{2} \pi c_1 h \frac{f \, v^2}{f'} \,\,\,
\label{eq:physmass}
\end{equation}
where \begin{equation}
v \equiv \sqrt{f^2+f^{\prime\,2}}=246 \ {\rm GeV}.\end{equation}
In the holographic treatment of this model, the parameter $c_1$ can be calculated, as well as the pion
couplings to the hadronic technicolor resonances in the theory.

\section{Holographic Technicolor with a Scalar}\label{sec:AdSCFT}

We use the AdS/CFT correspondence \cite{AdSCFT} to model the strong dynamics of the technicolor sector.
The holographic description allows us
to calculate the masses and couplings of the technicolor resonances and to
estimate the $S$ parameter.  We take the geometry of the 5D spacetime to be a slice of
anti-de Sitter (AdS) space, given by the metric,
\begin{equation}
ds^2 = \frac{1}{z^2}\left(-dz^2+dx^{\mu}dx_{\mu}\right), \qquad
0<z\leq z_m \label{eq:metric}
\end{equation}
where $z=z_m$ is an infrared cutoff.  We include in the 5D bulk a complex scalar field $X$, whose
boundary value is proportional to the source for the technifermion operator $q_{L} \bar{q}_R$ in the 4D
theory, where $q=(p,m)$.  The field $X$ is a two-by-two matrix in flavor space and transforms
as a bifundamental under the SU(2)$_L\times$SU(2)$_R$ chiral symmetry,
which becomes a gauge symmetry in the corresponding 5D model.
Normalizable modes of the $X$ field and the bulk gauge
fields correspond to hadronic resonances, with $1/z_m$ setting the
scale of confinement.  An ultraviolet cutoff may be introduced by moving the AdS boundary away
from $z=0$, to $z=\epsilon$.  One can think of $1/\epsilon$ as the scale at which the holographic
model breaks down.  Although we work with a finite and small choice for $\epsilon$ in our
numerical calculations, we find that all our physical
results remain convergent in the limit
$\epsilon \rightarrow 0$.  For definiteness, we also set the AdS scale to
the electroweak scale, $v=246\ {\rm GeV}$.

If we assign to the 4D operator $q_L \overline{q}_R$ its UV conformal dimension 3, then according
to the AdS/CFT correspondence, the corresponding 5D field $X$ has a mass squared $m^2_X=-3$ in units
of the AdS curvature scale~\cite{AdSCFT}. In principle, we can consider this mass, or equivalently
the dimension of the techniquark bilinear,  as another parameter in the model, and the running of
the dimension can be included by a modification of the geometry.  For simplicity we do not consider
such modifications here. To summarize the model and our conventions, the 5D action is,
\begin{equation}
S_{5D} = \int d^5x
\sqrt{g}\quad\mbox{Tr}\left\{-\frac{1}{2g_5^2}(F_R^2+F_L^2)+|DX|^2+3|X|^2\right\} \,\,\,,
\label{eq:5Daction}
\end{equation}
where $D_\mu X =\partial_\mu X-iA_{L\mu}X+i X A_{R\mu}$,
$A_{L,R}=A^a_{L,R}T^a$, and $F_{L,R\,\mu\nu}=\partial_\mu
A_{L,R\,\nu}-\partial_\nu A_{L,R\,\mu}-i[A_{L,R\,\mu},A_{L,R\,\nu}]$. The
profile of the $X$ field is determined by solving the classical
equations of motion with $A_{L,R}^\mu=0$ and $X(x,z)=X(z)$.  There
are two independent solutions, whose coefficients have the interpretation
of the common techniquark mass, $m_q=hf^\prime/\sqrt{2}$, and condensate,
$\sigma$, so that
\cite{AdSQCD2,AdSQCD3},\begin{equation}
X(z) =\frac{1}{2}\left( \frac{hf'}{\sqrt{2}} \, z + \sigma \, z^3\right) \,\,\,.\label{eq:v}
\end{equation}
In an SU($N$) technicolor theory with two flavors, one may match the
holographic prediction for the vector current-current correlator in the UV \cite{SVZ} to
the result of a one-loop calculation, which implies~\cite{AdSQCD2,AdSQCD3},
\begin{equation}
g_5^2 = \frac{24\pi^2}{N}\,\,\,. \label{eq:g5}
\end{equation}
Since we do not assume that $G_{TC}=\mbox{SU}(N)$ in the present model, however, $g_5$ is a
free parameter.  We will set $g_5$ to the value given by Eq.~(\ref{eq:g5}) with $N=4$ or $8$ for
the purpose of numerical estimates.  Our qualitative conclusions do not depend strongly on this choice.
Finally, for fixed small Yukawa coupling $h$
the condensate $\sigma$ can be expressed
as a function of the decay constant $f$ by a holographic calculation of the
small $q^2$ behavior of the axial vector current correlator
\cite{AdSQCD2,AdSQCD3}, $\Pi_A(-q^2)\rightarrow-f^2$.
Taking into account the constraint that $f^2+f^{\prime\,2}=v^2$, the free parameters in the
model are therefore $f$, $h$ and $z_m$.\footnote{The AdS/CFT correspondence
allows us to calculate the pion decay constant in terms of the techniquark
condensate in our model.  This provides a test of the naive dimensional
analysis (NDA) prediction, Eq.~(\ref{eq:condef}).  We find that $f(\sigma)$
agrees with NDA to within
a factor of ${\cal O}(1)$.  However, the discrepancy leads to qualitatively
different estimates of the $S$ parameter and decay widths.  In particular, the
$S$ parameter is generally smaller than our quoted
results if we take $\sigma=4\pi f^3$.}

The type of holographic construction that we have just described is known to give a reliable
description of the light pseudoscalar and vector mesons in QCD~\cite{AdSQCD2,AdSQCD3,AdSQCD1},
so we anticipate
that it will be equally successful in describing the technicolor sector of our 4D model.
We will use the correlation functions, masses and couplings computed in this theory to
determine unknown coefficients in the effective 4D chiral Lagrangian, described in
Sec.~\ref{sec:4DModel}, which properly takes into account the gauging of electroweak symmetry and
the mixing between the $\Phi$ and technipion fields. As we make explicit in the next section, our
approach does not require that we include the weakly coupled degrees of freedom in the 5D theory
to extract the desired results.

The 5D model that we have described is a simple holographic construction of the
strongly coupled sector, but is by no means the only one.  Additional interactions may be
included in the 5D action, the metric may be allowed to deviate from the AdS metric away from $z=0$,
and the boundary conditions for the fields at $z_m$ may be altered.
Such modifications make it possible to include power corrections to the
vector and axial-vector current-current correlation functions, so that one may obtain negative values
of the $S$ parameter~\cite{sanz}. Alternatively, $S$ may be reduced if the dimension of
the operator $\overline{q}q$ is smaller than its UV dimension \cite{hong}.
Although our model may be modified in these ways, we take a different approach. We work with the minimal
theory, Eqs.~(\ref{eq:metric}) and (\ref{eq:5Daction}), but allow the scale $z_m$ and the chiral symmetry
breaking scale $4 \pi f$ to be independent.  This freedom provides another means of obtaining a significant
reduction in $S$ (see also Ref.~\cite{Agashe}). With the confinement scale held fixed, we will also see that
$S$ decreases as the vev of the field $\Phi$ approaches the electroweak scale $v$, the limit in which the
technicolor sector no longer participates in EWSB.

\section{Phenomenology}\label{sec:Pheno}

In this section, we compute what is usually the most dangerous contribution to the $S$
parameter, and show that an acceptably small value can be obtained without adding new parameters
to the minimal 5D theory defined by Eqs.~(\ref{eq:metric}) and (\ref{eq:5Daction}).  We also consider
some aspects of the phenomenology of our holographic bosonic technicolor model that are
relevant to future collider searches.

\subsection{The $S$-Parameter}

One of the most serious problems with QCD-like technicolor models is the generically large
value of the $S$-parameter.  The oblique parameter $S$ may be
defined in terms of correlation functions of the vector and axial-vector
currents $J_V^{a\,\mu}$ and $J_A^{b\,\mu}$ at small momentum transfer~\cite{PT},
\begin{equation}
S=4\pi\,\frac{d}{dq^2}\left.\left(\Pi_V(-q^2)-\Pi_A(-q^2)\right)\right|_{
q^2\rightarrow 0},
\end{equation}
with, \begin{eqnarray}
\int d^4x e^{iq\cdot x}\langle J_V^{a\,\mu}(x)\,J_V^{b\,\nu}(0)\rangle&\equiv&
\delta^{ab}\,\left(\frac{q^\mu q^\nu}{q^2}-g^{\mu\nu}\right)\,\Pi_V(-q^2), \nonumber
\\
\int d^4x e^{iq\cdot x}\langle J_A^{a\,\mu}(x)\,J_A^{b\,\nu}(0)\rangle&\equiv&
\delta^{ab}\,\left(\frac{q^\mu q^\nu}{q^2}-g^{\mu\nu}\right)\,\Pi_A(-q^2).
\label{eq:Pi}\end{eqnarray}
In the holographic model, the contribution of the strong technicolor sector to
$\Pi_V(-q^2)$ and $\Pi_A(-q^2)$ are calculated by evaluating the part of the 5D
action, Eq.~(\ref{eq:5Daction}), that is quadratic in the SU(2)$_V$ and SU(2)$_A$
gauge fields.  According to the
rules of the AdS/CFT corresponence, the variation of the
action (twice) with respect to the 4D vector or axial vector gauge fields $V_\mu(q)$
or $A_\mu(q)$, which act as sources for $J_V^\mu$ and $J_A^\mu$,
yields the correlators in Eq.~(\ref{eq:Pi}).  We define the vector bulk-to-boundary
propagator $V(q,z)$ as the solution to the equations of motion for
the SU(2)$_V$ gauge field $V_\mu(q,z)\equiv V(q,z) V_\mu(q)$, where
$V(q,\epsilon)=1$ if $z=\epsilon$ is the location of the spacetime boundary;
similarly, the axial vector bulk-to-boundary propagator $A(q,z)$ is defined by
$A_\mu(q,z)\equiv A(q,z) A_\mu(q)$.
The desired correlators are then determined to be \cite{AdSCFT},
\begin{equation}
\Pi_V(-q^2)=\frac{2}{g_5^2}\left.\frac{1}{z}\frac{\partial V(q,z)}{\partial z}\right|_{z
\rightarrow \epsilon},
\end{equation}
and similarly for $\Pi_A(-q^2)$ with the replacement $V(q,z)\rightarrow A(q,z)$.
The equations of motion for the bulk-to-boundary propagators follow from the
action, Eq.~(\ref{eq:5Daction}): \begin{equation}
\partial_z
\left(\frac{1}{z}\partial_z V(q,z)\right) +\frac{q^2}{z}\,V(q,z) = 0,
\end{equation}
\begin{equation}
\partial_z\left(\frac{1}{z}\partial_z A(q,z)\right)
+\frac{q^2}{z}\,A(q,z)-\frac{g_5^2\,X_0(z)^2}{2z^3}\,A(q,z)=0.
\end{equation}

\begin{figure}[t]\centerline{
\epsfig{file=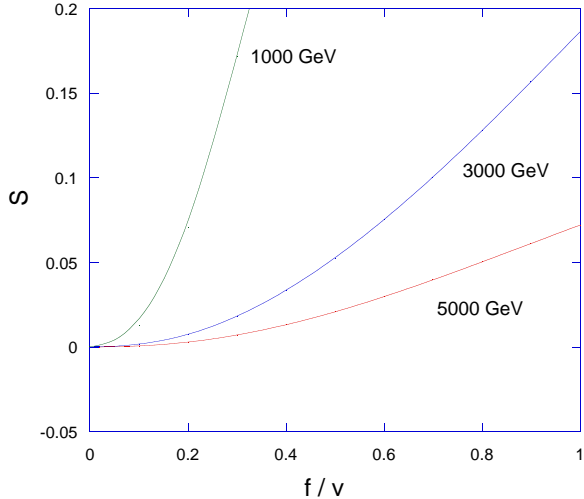, width=3in}}
\caption{Technicolor contribution to
$S$ parameter vs. technipion decay constant $f$ and lightest
technivector meson mass.  Curves for technivector masses of  $1$, $3$, and $5$
TeV are shown.  In this example the technifermion Yukawa coupling was taken to be $h=0.01$, and $g_5$
was chosen to match the UV behavior of the technicolor group SU(4) as described
in the text. Note that the existence of the Higgs field allows
$S<0.05$ over a significant region of the parameter space.}\label{fig:S}
\end{figure}

In our model, we allow for the possibility that confinement and chiral symmetry
breaking occur at different scales.  The confinement scale associated with
the masses of the vector mesons is determined by the shape of the
extra dimension away from the boundary at $z=\epsilon$.  In our model, the
location of the IR wall, at $z=z_m$, determines this scale.  By increasing
the confinement scale, physics around the $Z$ pole becomes increasingly
like the Standard Model, and corrections to $S$ become negligible\footnote{
Perturbative unitarity places an upper bound on the technirho mass.  However, for
heavier technirho, the 5D theory becomes strongly coupled and 5D loop effects that we have
ignored become important. We thank Csaba Cs\'{a}ki and Kaustubh Agashe for discussions of this issue.}.
Similarly, if the Higgs vev approaches the electroweak scale, $v$=246 GeV, with the technirho mass held fixed,
then the physics around the $Z$ pole again becomes increasingly
like the Standard Model.  (The technirho mass is calculated holographically, as will be discussed
in Sec.~\ref{sec:Pheno}B.) This behavior is reflected in Fig.~\ref{fig:S}.
The different curves correspond to different masses of the lightest technivector resonance, and hence
to different $z_m$. As $f/v\rightarrow 0$ the technicolor sector plays no role in EWSB, and as
$f/v\rightarrow 1$ the Higgs sector plays no role in EWSB.  Note that we take a reasonably small
(though not atypical) value for the Yukawa coupling $h=0.01$ in this example.  We have studied the
dependence of the $S$ parameter on $h$, for a technivector mass of 3 TeV, and have found that our results
remain unchanged at the few percent level for any $h\alt 0.3$.  We do not discuss larger $h$ since there
are then regions of the parameter space for which the approximation of chiral symmetry breaks down. The 5D
gauge coupling $g_5$ was taken to be as in Eq.~(\ref{eq:g5}) with $N=4$.  As we mentioned earlier, this
identification is made for definiteness but is somewhat arbitrary, as the UV description of the holographic
technicolor theory is unconstrained for our purposes.  It is clear from Fig.~\ref{fig:S} that without a large
hierarchy between the confinement and chiral symmetry breaking scales the $S$ parameter can be acceptably small.
This is unlike traditional technicolor models.

\subsection{Physical Spectrum and EWSB}

Thus far, the SU(2)$_L \times$SU(2)$_R$ chiral symmetry of the technicolor theory on
the 4D boundary has been assumed to be a global symmetry.  Strictly speaking, the
model described in Sec.~\ref{sec:AdSCFT} allows us to calculate hadronic properties in the limit
that the electroweak gauge interactions and couplings to the $\Phi$ scalars
are turned off.  In order to become a model of EWSB, we must now consider the effect of
gauging an SU(2)$\times$U(1) subgroup of the chiral symmetry.  As far as the strong technicolor
interactions are concerned, the only difference is that three Goldstone bosons are eaten
through the usual Higgs mechanism and are replaced by the longitudinal components of the $W$
and $Z$ bosons.  The 4D theory initially contains six pseudoscalar fields, the $\Pi_a$ and
$\Pi'_a$ defined in Sec.~\ref{sec:4DModel}, for $a=1\ldots 3$.  One linear combination, $\pi_a$, is
eaten during EWSB, and the other, $\pi_p$, remains in the physical spectrum, as given in
Eq.~(\ref{eq:piapip}).

The $\Pi^a$ components of $\pi_a$ and $\pi_p$ correspond to normal mode solutions of the bulk
equation of motion in the holographic theory, as we shall now review.
Ignoring its radial $\sigma$ component, we may express the bulk scalar field as
\begin{equation}
X(x,z)=\frac{X_0(z)}{2}\,\exp[2i\Pi_X(x,z)]  \,\,\,,
\end{equation}
where $X_0(z)=2 X(z)$, with $X(z)$ given in Eq.~(\ref{eq:v}). The quadratic part of the action,
Eq.~(\ref{eq:5Daction}), includes mixing between $\Pi_X$ and the longitudinal component of the axial
vector field $A_\mu(x,z)=\partial_\mu \varphi(x,z)$. The relevant terms in the action are: \begin{eqnarray}
S_{5D}&\supset& \int d^5x\,\left[-\frac{1}{4g_5^2\,z}\,F_{A\,MN}^aF_A^{MN\,a}+
\frac{1}{z^3}\left|D_MX\right|^2 \right]\\
&\supset& \int d^5x\,\left[-\frac{1}{4g_5^2\,z}\,F_{A\,MN}^aF_A^{MN\,a}+
\frac{X_0(z)^2}{2z^3}\left(\partial_M \Pi_X^a-
\frac{A_M^a}{\sqrt{2}}\right)^2\right].
\label{eq:s5d}
\end{eqnarray}
In the subsequent numerical analysis, we choose $g_5$ to match the UV behavior of the
technicolor group SU(8). For the normalizable modes, we write $\Pi_X(q,z)=\pi_X(z)\Pi(q)$ and
$\varphi(q,z)=\varphi(z)\Pi(q)$, where $q^2=m_\Pi^2$ is an eigenvalue of
the equations of motion with appropriate boundary conditions.
We use the gauge $A_z=0$, in which case the equations of motion for $\pi_X$ and
$\varphi$ are: \begin{equation}
\partial_z\left(\frac{1}{z}\partial_z\varphi^a\right)+\frac{g_5^2 X_0(z)^2}{
z^3\sqrt{2}}\left(\pi_X^a-\frac{\varphi^a}{\sqrt{2}}\right)=0,
\end{equation}\begin{equation}
-\frac{\sqrt{2}q^2}{g_5^2}\partial_z\left(\frac{1}{z}\partial_z\varphi^a\right)
+\partial_z\left(\frac{X_0(z)^2}{z^3}\partial_z\pi_X^a\right)=0.
\end{equation}
Note that the linearized
equations of motion are invariant under the gauge transformation
$\varphi^a/\sqrt{2}\rightarrow \varphi^a/\sqrt{2}+\lambda^a(q)$,
$\pi_X^a \rightarrow \pi_X^a+\lambda^a(q)$.  The boundary
conditions for the normalizable modes are,
$\pi_X^a(\epsilon)=\varphi^a(\epsilon)=0$, and $\left.\partial_z
\varphi^a(z)\right|_{z=z_m}=0$.
The UV boundary conditions are determined by the normalizability of the
modes (up to a gauge transformation as described above),
and the gauge invariant form of the IR boundary condition
is $F_{z\mu}=0$ (although other choices are possible).

\begin{figure}[t]\centerline{
\epsfig{file=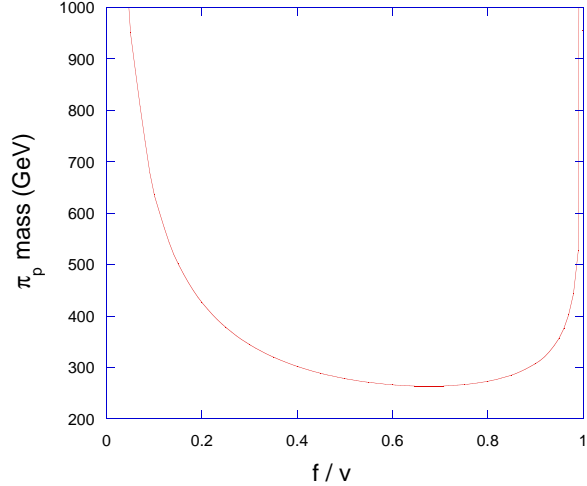, width=3in}}
\caption{The physical pion mass as a function of the technipion decay
constant $f$, for $h=0.01$ and $m_\rho=3$~TeV.  Given a generic potential for
the scalar doublet $\phi$ from Eq.~(\ref{eq:Phi}), one finds that the limit
$f/v\rightarrow 1$ is not physically accessible.}
\label{fig:mpi}
\end{figure}

The eigenvalues of $q^2$ for the coupled equations of motion determine
the mass squared term for the normalized $\Pi(x)$ field. In the holographic approach,
the $\Pi^2$ mass term in the 4D effective Lagrangian, arising from the
expansion of Eq.~(\ref{eq:keyterm}), is roughly proportional to the techniquark mass
and condensate (for small techniquark mass)
by the Gell-Mann--Oakes--Renner relation~\cite{GMO}, $m_\pi^2 f^2 \simeq
2 m_q \sigma$.
The techniquark mass and condensate appear in the profile of
the bulk scalar field $X(z)$ as in Eq.~(\ref{eq:v}), and indeed the
Gell-Mann--Oakes--Renner relation can be derived from the holographic model \cite{AdSQCD2}.
Taking into account the mixing in Eq.~(\ref{eq:piapip}), it follows that
\begin{equation}
m_\pi^2 \pi_p^a \pi_p^a = \frac{m_\pi^2}{v^2}\left[
f^{\prime\,2}\Pi^a\Pi^a-2 f^\prime f \Pi^a\Pi^{\prime\,a}+
f^2\Pi^{\prime\,a}\Pi^{\prime\,a}\right].
\end{equation}
The holographic calculation provides information on the $\Pi^2$ squared mass term
alone, $m_\Pi^2=  m_\pi^2\,f^{\prime\,2}/v^2$, with $v=246$~GeV.
This allows us to infer the physical pion mass in the full theory.  In Fig.~\ref{fig:mpi}, we plot
the physical technipion mass, $m_\pi$, as a function of $f/v$.

The analysis of the vector (technirho) and transverse axial vector sectors are more
straightforward in this model. The lightest axial vector resonance will be
heavier than the technirho, and of somewhat less interest in collider searches,
so we will not discuss it further here.  Considering only quadratic terms
in the 5D action, Eq.~(\ref{eq:5Daction}), the vector part of the SU(2)$_L \times$ SU(2)$_R$
gauge fields does not mix with either the axial part or with the bulk scalar $X$.
The equation of motion for the transverse part of the vector field in the
gauge $V_z^a=0$ is, \begin{equation}
\partial_z
\left(\frac{1}{z}\partial_z V_\mu^a(q,z)\right) +\frac{q^2}{z}\,V_\mu^a(q,z) = 0,
\end{equation}
with boundary conditions $V_\mu(q,\epsilon)=0$ and $F_{\mu z}(q,z_m)=0$.
The solutions are Bessel functions and the spectrum is given by zeroes of
$J_0(q\, z_m)$.  The mass of the lightest technirho in the model is therefore,
\begin{equation}
m_\rho= \frac{2.405}{z_m}.
\end{equation}

\subsection{Decays of the Technirho}

As long as phase space allows, the technirho will decay strongly about 100\% of
the time.  In our model
the dominant decays of the neutral technirho $\rho^0$
will be to the longitudinal $W$
boson and to physical pions $\pi_p$.  We will calculate the couplings
$g_{\rho\pi_p\pi_p}$, $g_{\rho W_LW_L}$ and $g_{\rho W_L\pi_p}$ that appear
in the effective 4D Lagrangian, \begin{equation}
{\cal L}_{\rho XY}= ig_{\rho XY}\,\rho_{0}^{\mu}\left[
(\partial_\mu X^+)Y^- - Y^+(\partial_\mu X^-)\right],
\label{eq:rXY}
\end{equation}
where $X$ and $Y$ represent either $W_L$ or $\pi_p$.
By the Goldstone boson equivalence theorem
\cite{equivalence-thm} we
treat the longitudinal $W$ as a Goldstone scalar (an unphysical
pion $\pi_a$) with mass $m_W$ coupled as in Eq.~(\ref{eq:rXY}).
The equivalence theorem is valid if the $W$ boson carries energy much
larger than its mass, which is valid in our examples for all of the
partial decays of the technirho except to $W_L \pi_p$
in the small region of parameter space for which $m_{\pi}\sim m_\rho$.
In that regime, however, the branching fraction
for decays to $W_L \pi_p$ is small anyway due to the reduced phase space, so
our qualitative results remain unchanged.

A standard calculation
of the partial decay widths gives, \begin{eqnarray}
\Gamma_{\pi_p\pi_p}&=&\frac{1}{48\pi} m_\rho \,
g_{\rho\pi_p\pi_p}^2
\left(1-4\frac{m_\pi^2}{m_\rho^2}\right)^{3/2} , \\
\Gamma_{W_L W_L}&=&\frac{1}{48\pi} m_\rho \,
g_{\rho W_L W_L}^2
\left(1-4\frac{m_W^2}{m_\rho^2}\right)^{3/2} ,\\
\Gamma_{W_L^+ \pi_p^-}=\Gamma_{W_L^- \pi_p^+}&=&\frac{1}{48\pi} m_\rho \,
g_{\rho W_L\pi_p}^2\left(1+\frac{m_\pi^4}{m_\rho^4}+\frac{m_W^4}{m_\rho^4}
-2\frac{m_W^2}{m_\rho^2}-2\frac{m_\pi^2}{m_\rho^2}-2 \frac{m_\pi^2 m_W^2}{
m_\rho^4}\right)^{3/2} .  \nonumber \\
&&\end{eqnarray}
In terms of the mixing angles $\cos\theta\equiv f/v$, $\sin\theta =
f^\prime/v$,
the couplings of the technirho are related by, \begin{equation}
g_{\rho \pi_p\pi_p} = g_{\rho W_L W_L}\,\tan^2 \theta  =
g_{\rho W_L\pi_p} \,\tan\theta
.\end{equation}

\begin{figure}
\begin{center}
$\begin{array}{cc}
%\multicolumn{1}{l}
%{\mbox{\bf }} &
%   \multicolumn{1}{l}{\mbox{\bf }} \\ [-0.53cm]
\epsfxsize=3in
\epsffile{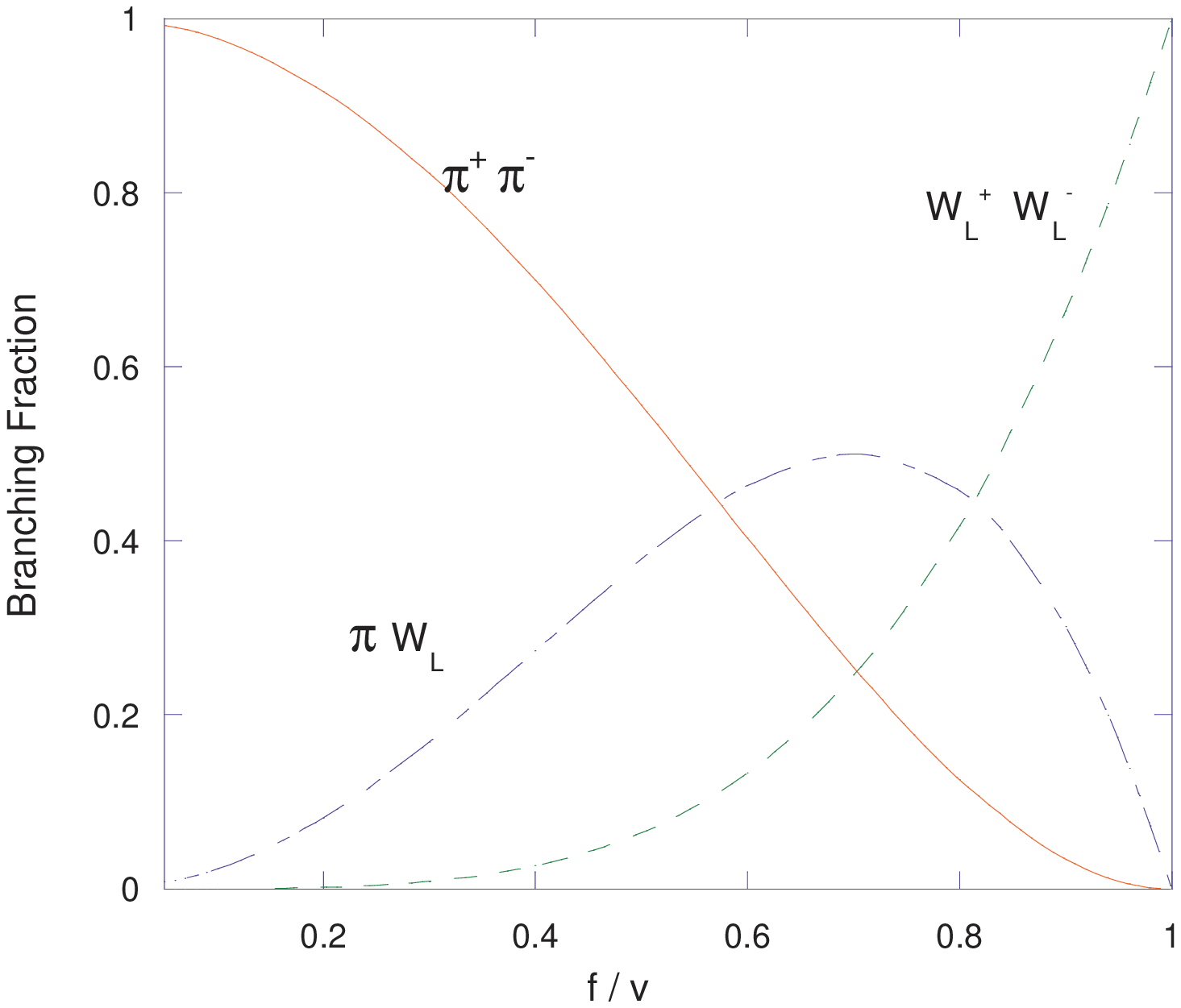} &
    \epsfxsize=3in
    \epsffile{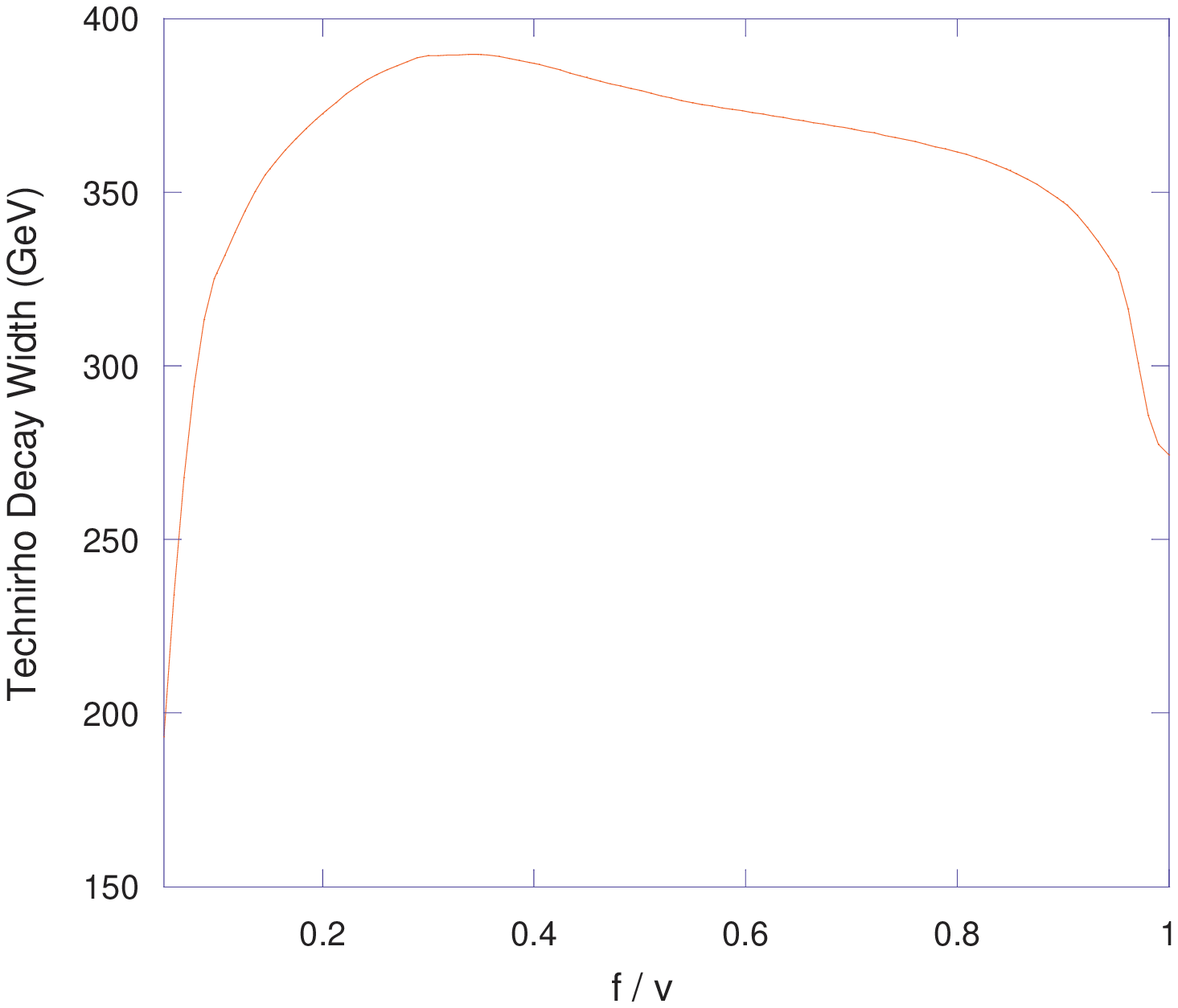} \\ [0.4cm]
%\mbox{(a)} & \mbox{(b)}
\end{array}$
\end{center}
\caption{Branching fractions and total decay width
of the $\rho^0$, for $m_\rho= 3$~TeV, and h=0.01.}
\label{fig:decays}
\end{figure}

Defining $\Gamma_{Tot}=\Gamma_{\pi_p^+ \pi_p^-} +
2 \Gamma_{W_L^+ \pi_p^-} + \Gamma_{W_L^+ W_L^-}$,
the branching fractions $\Gamma_{XY}/\Gamma_{Tot}$ depend
only on the mixing angles, resonance masses and $m_W$.
To calculate the total width we need to know $g_{\rho \pi_p \pi_p}$, which
is obtained in the holographic model by integrating the 5D action over
the extra dimension $z$ for the lightest modes of $V_\mu$ and $\Pi_X$, and
multiplying by the appropriate mixing angles to convert $\Pi$, defined after Eq.~(\ref{eq:s5d}),
to the physical pion $\pi_p$.  The couplings arise from both the gauge field
and scalar kinetic terms.  In terms of the modes $V_\mu(k,z)=
\psi_\rho(z)V_\mu(k)$,
$\varphi(q,z)=\varphi(z)\Pi(q)$ and $ \Pi_X(q,z) = \pi_X(z) \Pi(q)$, where
$k^2=m_\rho^2$ and $q^2=m_{\Pi}^2$ are the lowest
eigenvalues for the bulk equations of motion, we obtain the
$\rho\Pi\Pi$ coupling: \begin{equation}
g_{\rho \Pi \Pi}=\frac{g_5}{\sqrt{2}}\int dz\ \psi_\rho(z)
\left(\frac{\varphi'(z)^2}{
g_5^2\,z}+ \frac{X_0(z)^2\,\left(\pi_X(z)-\varphi(z)/\sqrt{2}\right)^2}{
z^3}\right) .
\label{eq:grpp}\end{equation}
The technirho wavefunction is
normalized such that $\int (dz/z) \psi_\rho(z)^2 =1$; the technipion
wavefunctions $\pi_X(z)$ and $\varphi(z)$ are
normalized such that the integral in
Eq.~(\ref{eq:grpp}) (without the prefactor) would
equal 1 if $\psi_\rho(z)$ were replaced by 1.  These normalizations are
chosen so that the modes are canonically normalized in the effective
4D theory \cite{AdSQCD2,AdSQCD3}.
The remaining pion fields $\Pi^\prime$ do not couple strongly
to the technirho in this model, so the contribution from the bulk 5D action completely
determines the coupling of the technirho to physical technipions.
Taking into account the mixing between the two sets of pions,
the physical pion coupling is then given by,
\begin{equation}
g_{\rho\pi_p\pi_p}=\sin^2\theta
\,g_{\rho\Pi\Pi} . \end{equation}
The branching fractions and total decay widths
are plotted in Fig.~\ref{fig:decays} as a function
of the mixing angle $\cos\theta$ for a fixed $z_m$ corresponding to
a technirho mass $m_\rho = 3$~TeV.  Note that the physical
pions become heavy as $\cos\theta\rightarrow 1$, so the branching fractions
to final states containing pions vanish in that limit for any $z_m$.

\section{Conclusions}\label{sec:Conclusions}
The condensation of a fermion bilinear operator could provide a simple mechanism for the
breaking of electroweak symmetry.  Nature provides
an example of this mechanism in QCD: a quark condensate spontaneously
breaks the global chiral symmetries of the theory, leading to the observed
spectrum of pseudogoldstone bosons.  Although the analogy to QCD is a source
of inspiration for theories of dynamical electroweak symmetry breaking, it
has also presented a challenge.  A technicolor sector that
is simply a scaled-up version of QCD produces a significant positive contribution
to the electroweak $S$ parameter, leading to results that are in conflict with current
experimental bounds.

There is no reason to believe {\em a priori}
that nature should choose a technicolor
sector that can be compared so easily to QCD.  Since the technicolor gauge coupling
is nonperturbative at the electroweak scale, however, alternative models have proved difficult
to study.  In the absence of simple scaling arguments that start with known
results from hadronic physics, one in the past could only make the polite observation that the $S$
parameter might not be a problem in all models.  The low-energy spectrum and
dynamics of any specific proposal could not be determined with any degree of certainty.

Holography presents a way around this impasse by allowing one to work instead
with an equivalent higher-dimensional theory that is weakly coupled.  We have
studied in this work a technicolor sector that is defined entirely in terms of its
five-dimensional holographic dual, allowing us to deviate in a calculable way from the
QCD-like limit.  We have shown that appropriate choices of the parameters in the 5D theory
exist where the otherwise leading contribution to the $S$ parameter is small.
Separation of the confining and chiral symmetry breaking scales in the
holographic approach is possible (although there is not necessarily a simple
4D gauge theory description of such a theory), and provides a mechanism for
reducing the $S$ parameter. We take this approach for simplicity, not out of necessity.
The main issue that we address is the generation of fermion masses. We have added to the theory a
(possibly composite) weak doublet field, whose vev is shifted from zero via its coupling to the
technifermion condensate.  Like a conventional Higgs doublet, the new field has Yukawa couplings
to the Standard Model fermions, allowing their masses to be generated.  Notably, the squared mass of
the scalar is taken to be positive in our model, so that it is not the origin of electroweak
symmetry breaking by itself.

The inclusion of this additional scalar doublet in our theory is not a radical
proposition.  It has been known for some time that models with  strongly-coupled extended technicolor
sectors can produce exactly this low-energy particle spectrum~\cite{setc}.  We have presented a holographic
representation of a strongly-coupled ETC theory in which low energy properties of the theory
are calculable.    Our model presents one possible mechanism for the generation of masses in holographic
technicolor models, an issue that has not yet been addressed in this context.

Using the AdS/CFT correspondence, we have calculated some basic quantities of phenomenological
interest in our model, namely the $S$ parameter, the lightest technirho and technipion masses,
and the dominant neutral technirho branching fractions. This provides a basis for future phenomenological
studies.   In particular, (i) a global electroweak analysis may provide useful constraints on
the Higgs boson and technirho masses, (ii) scalar-mediated flavor-changing-neutral-currents, which
were studied in similar effective theories~\cite{cg,cs1,cs2}, may also be re-examined in the present context, and (iii) the
production of technipions,  technivector and axial-vector resonances may be determined using
holographic estimates of the relevant couplings.  With the start of the LHC on the near horizon, a
detailed collider simulation that incorporates these results would be well motivated.

%%%%%%%%%%%%%%%%%%%%%%%%%%%%%%%%%%%%%%%%%%%%%%%%%%%%%%%%%%%%%%%%%%%%
\begin{acknowledgments}
We thank Kaustubh Agashe, Csaba Csaki, and Marc Sher for useful conversations. CDC
thanks the NSF for support under Grant No.~PHY-0456525. J.E. and J.T. thank the NSF for
support under Grant No.~PHY-0504442 and Jeffress Grant No.~J-768.
\end{acknowledgments}
%%%%%%%%%%%%%%%%%%%%%%%%%%%%%%%%%%%%%%%%%%%%%%%%%%%%%%%%%%%%%%%%%%%%

%\appendix
%\section{}
%\end{document}

%\bibliography{}

\end{document}